\documentclass[twocolumn,prl]{revtex4}

\usepackage{graphicx}
\usepackage{dcolumn}
\usepackage{bm}
\usepackage{amstext}

\begin{document}



\title{Theory of swimming filaments in viscoelastic media}
\author{Henry C. Fu}
\affiliation{Division of Engineering, Brown University, Providence, RI 02912}
\author{Thomas R. Powers}
\affiliation{Division of Engineering, Brown University, Providence, RI 02912}
\author{Charles W. Wolgemuth}%
\affiliation{Department of Cell Biology, University of Connecticut Health Center, Farmington, CT 06030}%

\date{\today}


\begin{abstract}

Motivated by the swimming of sperm in the non-Newtonian fluids of the
female mammalian reproductive tract, we examine the swimming of
filaments in the nonlinear viscoelastic Upper Convected Maxwell model.   We obtain
the swimming velocity and hydrodynamic force exerted on an
infinitely long cylinder with prescribed beating pattern.  We use
these results to examine the swimming of a simplified sliding-filament
model for a sperm flagellum.  Viscoelasticity tends to decrease 
swimming speed, and changes in the beating patterns due to
viscoelasticity can reverse swimming direction.

\end{abstract}


\maketitle


The physical environment of the cell places severe constraints on mechanisms for motility. For example, viscous effects dominate inertial effects in water at the scale of a few microns. Therefore, swimming cells use viscous resistance to move, since mechanisms that rely on imparting momentum to the surrounding fluid, such as waving a rigid oar, do not work~\cite{Ludwig1930,purcell1977}.  The fundamental principles of swimming in the low-Reynolds number regime of small-scale, slow flows have been established for many 
years~\cite{taylor1951,gi_taylor1952,lighthill1975,purcell1977}, 
yet continue to be an area of active research. 
However, when a sperm cell moves through the viscoelastic mucus
of the female mammalian reproductive tract, the theory of swimming in
a purely viscous fluid is inapplicable.  Observations of sperm show that
they are strongly affected by differences between viscoelastic
and viscous fluids.  In particular, the shape of the flagellar beating
pattern as well as swimming trajectories and velocities depend on the
properties of the medium \cite{HoSuarez2003, Suarez_et_al1991, SuarezDai1992}.  

The interplay of medium properties and flagellar motility or transport also arises in other situations, such as spirochetes swimming in a gel~\cite{CharonGoldstein2002}, and cilia beating in mucus to clear foreign particles in the human airway~\cite{SatirChristensen2007}. Motivated by these phenomena, we develop a theory for swimming filaments in a viscoelastic medium.  We begin by analyzing the swimming of an infinite filament with a prescribed
beating pattern in a fluid described by the Upper Convected Maxwell (UCM) model~\cite{BirdArmstrongHassager1977}. 
We deduce the hydrodynamic
force per unit length acting on the filament and the swimming
velocity to leading order in the deformation of the filament.  Our results extend the findings of Lauga~\cite{Lauga2007}, who considered a variety of fading memory models for the case of a prescribed beat pattern on a planar sheet.  We further apply our results to a simple model flagellum with active internal forces, and find that changes in flagellum
shapes play a crucial role in distinguishing the effects of
viscoelastic media.  



\begin{figure}
\includegraphics[width=8.6cm]{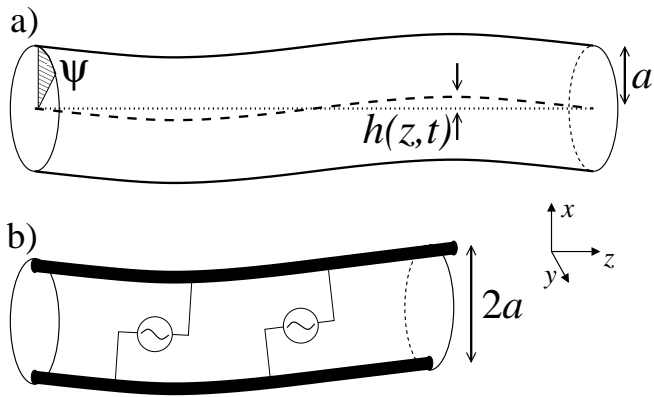}
\caption{a) Cylinder with imposed
  traveling wave of transverse displacements in a viscoelastic fluid.
b) Active flagellum model.  The active elements inside
  the flagellum represent motors that slide the filaments relative
  to each other.} \label{TaylorCylinder}
\end{figure}

Newtonian fluids  are characterized by a simple constitutive relation, in which stress is proportional to strain rate. Non-newtonian fluids cannot be characterized by a simple universal constitutive relation, and exhibit a range of phenomena such as elasticity, shear thinning, and yield stress behavior. We choose to focus our attention on  fluids with fading memory, in which the stress relaxes over time to the viscous stress. We consider small amplitude deflections of an infinite filament of radius $a$ (Fig. \ref{TaylorCylinder}a). Since the swimming velocity of a filament is second order in the deflection 
amplitude~\cite{taylor1951,gi_taylor1952}, linear models for fading memory such as the Maxwell model are insufficient for studying swimming~\cite{Lauga2007}.

Therefore, we use the simplest nonlinear constitutive relation incorporating elastic effects, the UCM model. This model is appropriate for a polymer solution in which the viscosity of the Newtonian solvent is disregarded:
\begin{equation}
{\bm \tau} + \lambda \hat{\bm  \tau} = \eta \dot {\bm \gamma}.\label{ucm}
\end{equation}
In this equation $\bm \tau$ is the deviatoric stress, $\lambda$ is the relaxation time, $ \hat {\bm \tau} =
\partial_t {\bm \tau} + \mathbf v \! \cdot \! \nabla \bm \tau -
{(\nabla \mathbf v)}^{\mathsf{T}} \! \! \cdot \! \bm \tau - \bm \tau \! \cdot \!
\nabla \mathbf v$ is the upper-convected time derivative of ${\bm \tau}$, $\mathbf{v}$ is the velocity, $\eta$ is the polymer viscosity, and
$\dot{\bm\gamma}=\nabla{\mathbf v}+{(\nabla \mathbf v)}^{\mathsf{T}}$
is the strain rate. The nonlinear terms of the upper-convected derivative make the constitutive relation insensitive to translational and rotational motion of material elements~\cite{BirdArmstrongHassager1977}.
The UCM fluid
responds as an elastic solid when subject to a rapidly varying
stress, and as a viscous liquid when subject to a slowly varying
stress.  When $\lambda=0$, the constitutive relation (\ref{ucm}) is Newtonian. 
Since inertia is unimportant, the motion of the  medium is governed by force balance, $-\nabla p +\nabla\cdot{\bm \tau}=0$, or
\begin{equation}
- (1 + \lambda \partial_t) \nabla p + \eta  \nabla^2\mathbf{v}
 = \lambda \nabla \cdot {\mathbf T},\label{eom1}
\end{equation} 
where $\mathbf T=\mathbf v \! \cdot \! \nabla \bm \tau -
{(\nabla \mathbf v)}^{\mathsf{T}} \! \! \cdot \! \bm \tau - \bm \tau \! \cdot \!
\nabla \mathbf v$,  and we have assumed incompressibility, $\nabla\cdot\mathbf{v}=0$.

To calculate the swimming velocity of the filament, we prescribe a beating pattern that is independent of load, and solve (\ref{eom1}) for the flow, imposing no-slip boundary conditions at the surface of the filament. Material points on the the surface of the filament are parameterized by $z$ and $\psi$  as in Fig.  \ref{TaylorCylinder}a. The positions of points on the surface are given by 
 \begin{equation}
{\mathbf r}(z,\psi,t) =  [h(z,t) + a \cos(\psi)] \hat
 {\bf x} + a \sin(\psi) \hat {\bf y} + z \hat {\bf z},    
\end{equation}
where $h(z,t) = \mathrm{Re} \sum_{q,\omega}  h_{qw} \exp(i q z
- i \omega t)) $. Note that we work in the frame in which material points of the cylinder move in planes of constant $z$. We will find that the no-slip boundary conditions can only be satisfied if there is a uniform flow along the $z$ axis. In the lab frame in which the fluid is at rest at infinity, the flow corresponds to the swimming velocity of the cylinder.


Invoking small amplitude, long wavelength distortions for simplicity, we solve the equations order by order in the 
displacement amplitudes $q h_{q\omega}$, to lowest order in $1/\log qa$,
and assuming that $h_{q\omega}/a \ll1$.  Expressing the dynamic variables as expansions in $q h_{q\omega}$, such as 
$\mathbf{v}=\mathbf{v}^{(1)}+\mathbf{v}^{(2)}+ ...$,
the no-slip boundary conditions in cylindrical coordinates $(r,\phi,z)$ up to
second order are
\begin{eqnarray}
\dot h \hat {\bf x}  &=& 
{\mathbf v}^{(1)} + {\mathbf v}^{(2)} +  h \left(
\cos \psi \; \hat {\bf r}  - \sin \psi \; \hat {\bm \phi} \right) \cdot \nabla {\mathbf
  v}^{(1)}
\end{eqnarray}
where $h$ is evaluated at $(z,t)$; and $\mathbf v$, $\hat\mathbf{r}$, and $\hat{\bm\phi}$ at $(a,\psi,z,t)$.
We used the fact that the azimuthal angle $\phi$ of the material point labeled $\psi$ on the cylinder surface is given by  $\phi\approx\psi-(h/a)\sin\psi$, and the radius of this point is given by $r\approx a+h\cos\psi$.
There is no first order contribution to $\mathbf
T$ and the first order dynamical equations are 
\begin{eqnarray}
(1 + \lambda \, \partial_t) {\bm \tau}^{(1)} &=& \eta \dot {\bm
  \gamma}^{(1)}  \label{firststress}\\
(1 + \lambda \, \partial_t) \nabla {p}^{(1)}
 &=& \eta  \nabla^2 {\bf v}^{(1)} \label{first},
\end{eqnarray}
with $\nabla\cdot\mathbf{v}^{(1)}=0$.
Since Eq.~(\ref{first}) is Stokes equation with a modified pressure,
the solution is readily found~\cite{happel_brenner1965}, and  
the first order flow $\mathbf{v}^{(1)}$ is the same as 
in the purely viscous case, in the limit
$qa \ll 1$.   The formulas for this flow are
given by Taylor in~\cite{gi_taylor1952}.   Just as in Taylor's case, the swimming velocity vanishes to first order in $qh_{q\omega}$. 
Using the pressure and stress, we find that the force per unit
length is purely in the $\hat {\bf x}$ direction:
\begin{equation}
\mathbf{f}_\mathrm{fluid}^{(1)}(z,t) = \mathrm{Re} \sum_{q,\omega} \frac{-4 \pi \eta i
  \omega h_{q\omega} }{(1 - i \lambda \omega)\log q a} \mathrm{e}^{i q z - i \omega t}\hat{\bf x}
\label{force}
\end{equation}
This expression is consistent with the results of Fulford, Katz, and Powell, who used resistive force theory for a filament beating in a linear Maxwell fluid to show that there is no change in the swimming velocity relative to the viscous case, even to \textit{second order} in $qh_{q\omega}$~\cite{FulfordKatzPowell1998}.  

In our problem, however, the nonlinearities make the viscoelastic swimming velocity different from the viscous swimming velocity. To second order, 
\begin{eqnarray}
(1 + \lambda \partial_t) {\bm \tau}^{(2)} &=& \eta \dot {\bm
  \gamma}^{(2)} - \lambda  {\mathbf T}^{(2)} \\
(1 + \lambda \partial_t) \nabla {p}^{(2)}
 &=& \eta \nabla^2 {\bf v}^{(2)} - \lambda \nabla
  \cdot {\mathbf T}^{(2)}  \label{second},
\end{eqnarray}
where $\mathbf T^{(2)}$  can be calculated using only
the first order stresses and velocity fields.  To find the time-averaged swimming velocity, we need only examine the velocity fields averaged over time and $\phi$. To second order, we find that there is a uniform flow at infinity, corresponding to a swimming velocity in the lab frame of
\begin{equation}
\mathbf{U}^{(2)} = -\frac{ 1}{2} \sum_{q,\omega}  \frac{|h_{q\omega}|^2 q \omega
  }{1 +
  (\lambda \omega)^2}  \hat{\mathbf z}.
\label{speed}
\end{equation}
%
For a single traveling wave the direction of swimming is opposite the
direction of motion of the traveling wave. Our result, valid to first order in an expansion in $1/\log(qa)$, is precisely the same as the case of a traveling wave on a planar sheet~\cite{Lauga2007}.

We have dealt with the effects of viscoelasticity on
a swimming filament with prescribed shape $h(z,t)$. 
In contrast, the beating patterns and the swimming
velocity of real sperm are affected by the medium, suggesting that prescribing the shape changes
of the swimmer may miss important effects.   Therefore, we consider a sliding filament model in which we prescribe active internal bending forces~\cite{brokaw1971,camalet_et_al1999}, and solve for the flagellum shape as well as the swimming speed. Our task is greatly simplified since the shape is required to an accuracy of first order in deflection only.
The model is shown in Fig.~\ref{TaylorCylinder}b. We continue to
assume the flagellum has a cylindrical cross-section as in
Fig~\ref{TaylorCylinder}a and deflection from the z-axis $h \hat
{\mathbf x}$, but now we assume the flagellum has finite length $L$ and two inextensible longitudinal filaments with constant lateral spacing $2a$. Motors along the flagellum attach to both filaments and tend to slide them past each other. Sliding is prohibited at the end near the head, which is omitted for simplicity. Assuming a planar shape of the flagellum, the moment $\mathbf{M}=M\hat\mathbf y$ acting at a cross-section of the flagellum consists of a passive resistance to bending, and an active part due to the sliding motors. To first order in deflection $h$, 
\begin{equation}
M(z)=A h''-2a\int^L_z f_\mathrm{m}\mathrm{d}z,
\label{activemoment}
\end{equation}
where $A$ is the bending stiffness, primes denote derivatives with respect to $z$, and $f_\mathrm{m}$ is the force per unit length that the lower filament of  Fig.~\ref{TaylorCylinder}b exerts on the upper filament. For simplicity we disregard  any elastic or viscous effects arising from proteins linking the filaments.

The balance of internal forces and hydrodynamic forces determines the instantaneous shape of the flagellum.  To find the internal force per unit length $f_\mathrm{int}$, consider moment balance on an element of the flagellum to first order in deflection~\cite{landau_lifshitz_elas}: $M'+N=0,$ where $N$ is the shear force acting on a cross section in the $x$ direction. Using $f_\mathrm{int}=N'$ yields 
$f_\mathrm{int}=-Ah''''-2af'_\mathrm{m}$. 
We choose a sliding force $f_m = \mathrm{Re} [f \exp( ik z -i \omega t)]$. To linear order, the shape change occurs with only one frequency $\omega$, and
$h(z,t) = \mathrm{Re}[ \tilde h(z) \exp(-i \omega t)]$.


Since the internal forces $f_\mathrm{int}$ are expressed in real space, it is convenient to write the hydrodynamic force Eq.~(\ref{force}) in real space. The logarithm in Eq.~(\ref{force}) varies slowly with $q$ and we replace it with a constant, $\log(qa)\approx\log(a/L)$, as is commonly done in resistive force theory. Thus
\begin{equation}
{\mathbf f}_{\mathrm{fluid}} = \mathrm{Re} \frac{ -
  \zeta_\perp }{1 - i \lambda \omega}  \left(-i \omega \tilde h(z) e^{-i \omega t} \right) \hat {\mathbf x},
\end{equation}
where $\zeta_\perp \approx 4 \pi \eta/
   \log\left( L/a\right) $.  

We non-dimensionalize our equations of motion
by measuring lengths in terms of $L$, $f_{\mathrm m}$ in
terms of $A/(2a L^2)$, and time in terms of $\omega^{-1}$.  For notational
simplicity, after scaling we use the same symbols for the new quantities.
The equation of motion for the active flagellum is $\mathbf{f}_\mathrm{int}+\mathbf{f}_\mathrm{fluid}=0$, or in non-dimensional form,
\begin{equation}
\frac{-i {\mathrm{Sp}^4 } }{1 - i
  \mathrm{De}} \tilde h + \tilde h'''' - ikf = 0.\label{NDEOM}\\
\end{equation}
The dimensionless ``Sperm number" ${\mathrm{Sp}}= L (\omega \zeta_\perp /
A)^{1/4}$ is the fourth root of the ratio of the
bending relaxation time of the flagellum to the period of the
traveling wave, and the Deborah number $\mathrm{De} = \lambda \omega$ measures
the importance of elastic effects.  Following~\cite{camalet_et_al1999}, we estimate $L=40$\,$\mu$m,
$a=20$\,nm, $A=4\times10^{-22}$\,N-m$^2$,
$\omega/(2\pi)=30$\,s$^{-1}$, and
$\zeta_\perp=2\times10^{-3}$\,N-s-m$^{-2}$, and a dimensional
magnitude of internal sliding forces $f_\mathrm{m}=4$\,pN/(24\,nm).
Therefore, $\mathrm{Sp} \approx 7$, and the 
dimensionless amplitude $f\approx13$.
For sperm which oscillate at frequency 25--50\,Hz in cervical mucus with
time constant $\lambda\approx1$--$10$\,s~\cite{TamKatzBerger1980}, 
we take
$\mathrm{De} \approx 100$.
Equation~(\ref{NDEOM}) must be supplemented by boundary conditions. For simplicity we forbid transverse motion of the head $ h(0)=0$, and suppose the connection between the head and the flagellum cannot support a moment: $ h''(0) + \int_0^1 f_{\mathrm{m}}(z) \mathrm{d}z = 0$ (see~\cite{LaugaFloppy2007} for a more realistic treatment of the moment boundary condition). The boundary condition at the other end is zero force, $ -  h'''(1) + f_{\mathrm m}(1) =
0$, and zero moment, $ h''(1) = 0$.  

\begin{figure*}
\includegraphics[width = 17.6cm]{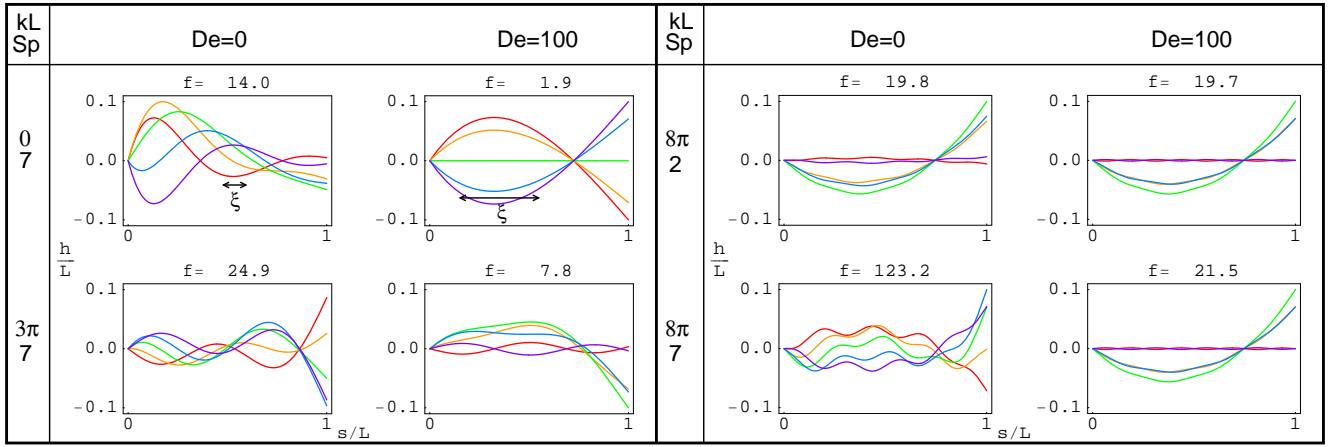}
\caption{ (color). Dimensionless amplitude $h/L$ versus dimensionless
  arclength $s/L$ for beating flagella with internal sliding forces.  A half-cycle
  (red, orange, green, light blue, blue) of the pattern is shown for
  viscous ($\mathrm{De}=0$) and viscoelastic ($\mathrm{De}=100$)
  cases.   For
  $\mathrm{Sp} =7$,  we show beating patterns for internal sliding
  forces with $k$ varying from $0$ (uniform force) to $8 \pi$. 
  We also show how changing the sperm number affects the
  shape for $k = 8 \pi/L$.  On the plots for $k=0$, we show $\xi$. 
  At the
  top of each plot, the (dimensionless) magnitude $f$ required to produce motion with amplitude $0.1 L$ is shown.} \label{shapesfig}
\end{figure*}

Representative plots of the beating patterns are shown in
Fig.~\ref{shapesfig}.   
Two lengthscales are apparent.  The first  is the wavelength of the sliding
force, $2 \pi/k$, readily apparent in the change of beating
patterns from $k=0$ to $k=8 \pi/L$ at $\mathrm{Sp} = 7$.  The
second lengthscale $\xi$ arises from the interaction of bending and
hydrodynamic forces: 
$\xi/L = |1 - i \mathrm{De}|^{1/4} / \mathrm{Sp}$.
The flagellum behaves like a rigid rod for large $\xi/L$, and is floppy for small $\xi/L$. Changing the viscoelastic properties of the fluid affects the beating
shapes through $\xi$, as  can be seen by
comparing beating patterns for $\mathrm{De}=0$ and $\mathrm{De} = 100$
in the flagellum with $k=8 \pi/L$ and $\mathrm{Sp} = 7$.  
In Fig.~\ref{shapesfig}, the amplitude of the sliding force is selected so that the maximum displacement of the flagellum is $L/10$.   Since $\xi$ increases with $\mathrm{De}$, smaller driving forces are required to produce the same amplitude motion.  Besides affecting $\xi$, viscoelasticity
affects phase behavior:  for $\mathrm{De}=0$, we
obtain traveling wave beating patterns, while for large $\mathrm{De}$, in
the elastic limit, we obtain standing wave beating patterns.  

By inserting the beating pattern $h_{q\omega}$ into Eq.~(\ref{speed}), we can calculate the swimming speed of a flagellum with prescribed internal forces. 
Although our flagellum is finite, we
may apply Eq.~(\ref{speed}) to our calculated 
beating
pattern 
by periodically  replicating it to infinite extent in $z$.  Physically, we are
ignoring end effects.  
It is
useful to rewrite Eq.~(\ref{speed}) in real space, for an infinite flagellum
with period $L$ and single oscillation frequency $\omega$:
\begin{equation}
\mathbf{U}^{(2)} = - \frac{1}{2 L (1 + \mathrm{De}^2)} \int_0^L \langle \dot h(z,t)
h'(z,t) \rangle \mathrm{d}z \; \hat {\mathbf z} 
\label{velocity}
\end{equation}
where the dot denotes the time derivative and the brackets denote time-averaging.  
The swimming velocity is independent of whether we consider the
period $L$ or $nL$; 
thus, the periodically replicated
flagellum has the same velocity as the single flagellum, up to end effects.

\begin{figure}
\includegraphics[width=8.6cm]{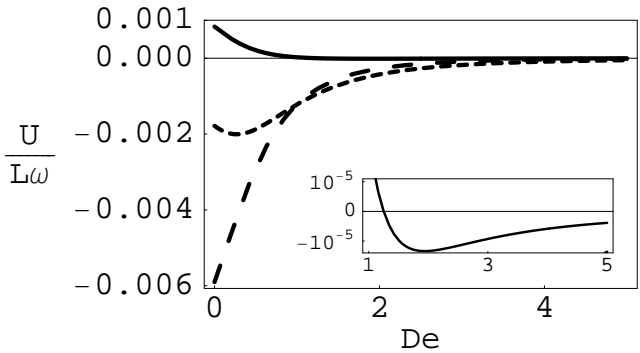}
\caption{Dimensionless swimming velocities versus De, for  $\mathrm{Sp} =
  7$ and maximum flagellum displacement $0.1 L$,
  for $k=0$ (long dashes), $k=3
  \pi/L$ (short dashes), $k= 8 \pi/L$ (solid).  The inset shows detail
  for the same plot demonstrating that viscoelastic effects can
  reverse the direction of swimming.} \label{velocityfig}
\end{figure}

The velocities of active flagella with sliding forces of varying
wavelength are plotted in Fig.~\ref{velocityfig}.   Viscoelastic effects can produce qualitative changes in
the swimming velocity.  Most notably,
as $\mathrm{De}$ is increased, the velocity can change direction, for
example in the case of $k= 8\pi /L$.  
When $k = 8
\pi/L$ and $\lambda=0$, there is a small amplitude traveling wave
moving in the positive $z$-direction with wavelength $\approx L/4$
driven by the sliding forces, but there is also a larger amplitude
traveling wave moving in the negative direction (Fig. \ref{shapesfig}).  A filament swims in the direction opposite to traveling waves moving in the
filament so these push the flagellum in opposite
directions, but the net velocity is in the positive direction.  As
$\mathrm{De}$ increases, the larger amplitude wave becomes a standing
wave, and only the smaller amplitude traveling waves remain, causing
the flagellum to move in the negative direction.  We emphasize that
this effect is not due to the $ (1+\mathrm{De}^2)$ correction factor to
the swimming speed in Eq.~(\ref{velocity}), which can never reverse the swimming direction. 
For swimming in viscoelastic fluids it is
crucial to allow the flagellum shape to respond to changed forces
rather than prescribe fixed beating patterns. 
We conclude by indicating directions for future work.  More realistic
constitutive relations should be explored, since the UCM model is
invalid for sufficiently high extensional flows and does not include
shear thinning effects.  Realistic modeling should also include
the effects of ends and large displacements disregarded here. Finally, it is important to determine whether the forces
imposed on the filaments by motors are the same in viscoelastic and
viscous media, or, instead, sperm respond to changes in load.
We note that the observation that beating frequency is dependent
upon the medium~\cite{HoSuarez2003, Suarez_et_al1991, SuarezDai1992} suggests the latter.

We thank A. Bower and E. Lauga for helpful conversations.  This work
is supported in part by National Science Foundation Grants NIRT-0404031 and DMS-0615919 (TRP); and NIH R01 GM072004 and
NSF CTS 0623870 (CWW).  We all thank the Aspen Center for Physics, and  HCF and TRP also thank the Hatsopoulos Microfluids Laboratory at MIT, where some of this work was completed.

\end{document}